\input{aipcheck}

\documentclass[
    ,final            
  ]
  {aipproc}

\layoutstyle{6x9}

\newcommand{\be}{\begin{eqnarray}}
\newcommand{\ee}{\end{eqnarray}}

\begin{document}
\quad \quad \quad \quad \quad \quad \quad \quad \quad \quad \quad \quad \quad \quad \quad \quad CPHT-PC098.1110, LPT-ORSAY  10-82
\title{Baryon to meson transition distribution
amplitudes and their spectral representation}


\author{B.~Pire}{
  address={CPhT, \'{E}cole Polytechnique, CNRS,  91128, Palaiseau, France}
}

\author{K.~Semenov-Tian-Shansky}{
  address={CPhT, \'{E}cole Polytechnique, CNRS,  91128, Palaiseau, France}
  ,altaddress={LPT,   Universit\'{e} d'Orsay, CNRS, 91404 Orsay, France} 
}

\author{L.~Szymanowski}{
  address={Soltan Institute for Nuclear Studies, Warsaw,  Poland.}
}

\begin{abstract}
 We consider the problem of construction of a spectral representation
 for nucleon to meson transition distribution amplitudes (TDAs), non-diagonal
 matrix elements of nonlocal three quark operators between a nucleon and
 a meson states. We introduce the notion of quadruple distributions and generalize
Radyshkin's factorized Ansatz for this issue. Modelling of baryon to meson TDAs
 in the complete domain of their definition opens the way to quantitative estimates of
 cross-sections for various hard exclusive reactions.
\end{abstract}

\maketitle

\section{Introduction}

Nucleon to meson transition distribution amplitudes (TDAs)
\cite{Frankfurt:1999fp,Frankfurt:2002kz,Pire:2005ax, Lansberg:2007ec}
(also known as super skewed parton DAs)
occur in the description of
backward meson%
\footnote{For definiteness we consider the pion case although the  analysis is completely general.}
electroproduction
$\gamma^*(q) + N(p_1) \rightarrow N(p_2) + \pi (p_{\pi})$
(as well as
$N(p_1)+\bar{N}(p_2) \rightarrow \gamma^*(q) + \pi (p_{\pi})$
in the near forward or backward regions)
in the framework of the collinear factorization approach.
Nucleon to meson TDAs represent new nonperturbative objects which
extend the concept of generalized parton distributions (GPDs).
Up to this moment  nucleon to meson TDAs lacked any suitable
phenomenological parametrization properly taking into account
the requirement of Lorentz covariance which is manifest as
the polynomiality property of the Mellin moment of TDAs in the
relevant light-cone momentum fractions, as well as the support properties.
For the case of GPDs this
kind of requirements can be fulfilled employing the spectral representation
in terms of double distributions (DDs) \cite{RDDA1}. The corresponding spectral
properties were established using the $\alpha$-representation techniques \cite{Radyushkin:1983ea}.
The factorized Ansatz for DDs
\cite{RDDA4} became the basis for a variety  of successful
GPD models.

Following the line of
\cite{Pire:2010if} we address the problem of constructing a spectral representation for
the case of nucleon to meson TDAs. This requires introduction of the notion of quadruple
distributions. We also generalize  Radyushkin's factorized Ansatz for these quantities.
We are thus ready  for phenomenological modelling of meson to nucleon
TDAs and quantitative estimates of cross-sections of new classes of hard exclusive reactions.

\section{Definition and properties of $\pi N$ TDAs}
$\pi N$
TDAs are defined as the Fourier transform
of the matrix element  between a nucleon and a pion states of the three-local light-cone quark operator.
In the light-cone gauge
$A^+=0$
their definition can be symbolically written as
\cite{Frankfurt:1999fp,Frankfurt:2002kz}:
\be
&&
\! \! \! \! \!
\int  \left[ \prod_{i=1}^3 \frac{dz^-_i}{2 \pi}  e^{ix_i(P \cdot z_i)} \right]
\left.
\langle \pi(P+\Delta/2) |
\epsilon_{abc}
\psi_{j_1}^a(z_1)
\psi_{j_2}^b(z_2)
\psi_{j_3}^c(z_3)
|N(P-\Delta/2) \rangle
\right|_{z_i^+=z_i^\bot=0}
 \nonumber \\ &&
 \! \! \! \! \!
\sim \delta(2 \xi-x_1-x_2-x_3) H_{j_1 \, j_2 \, j_3} (x_1,\,x_2,\,x_3,\, \xi, \, u)\,.
 \label{TDA_definition}
\ee
Here $j_{i}$ stand for spin-flavor indices and $a$, $b$, $c$ are color indices.
The tensor decomposition  of the r.h.s. of
(\ref{TDA_definition})
involves a set of independent spin-flavor structures multiplied by corresponding invariant functions
($\pi N$
TDAs). Each invariant function $H$ depends on three longitudinal momentum fractions
$x_i$ ($i=1,2,3$) subject to the constraint
$x_1+x_2+x_3=2 \xi$,
skewness parameter
$\xi=- (\Delta \cdot n)/( 2 P \cdot n) \, (n^2=0)$
and the momentum transfer squared
$u= \Delta^2$.

The support
property in $x_i$ is given by
$-1+\xi \le x_i \le 1+\xi$.
One can distinguish Efremov-Radyushkin-Brodsky-Lepage (ERBL)-like domain in which
all three momentum fractions $x_i$ are positive and
Dokshitzer-Gribov-Lipatov-Altarelli-Parisi (DGLAP)-like domains in which one or two
momentum fractions $x_i$ are negative.
It is convenient to introduce the so-called quark-diquark coordinates. There are three possible choices
depending on which quarks are supposed to form a ``diquark system''. Below we employ the particular
choice of quark-diquark coordinates:
$ v \equiv v_{3}=\frac{x_1-x_2}{2}$;  $w\equiv w_{3}=\frac{x_3-x_1-x_2}{2}$. We also introduce the notation
$ \xi' \equiv \xi'_3 = \frac{\xi-w_3}{2}$.
The complete domain of definition of
$\pi N$ TDA
in quark-diquark coordinates
can be parameterized as follows:
$-1 \le w \le 1$; $ -1+|\xi-\xi'| \le v \le 1- |\xi - \xi'|$.

One may check that the $(n_1,n_2,n_3)$-th
Mellin moment of $\pi N$ TDAs (\ref{TDA_definition}) is related to the form factors of local three quark operators
between nucleon and pion states and is a polynomial of $\xi$ of order $n_1+n_2+n_3$
\footnote{The problem of
necessity of adding  $D$-term like
contributions to $\pi N$ TDAs still lacks some analysis.
}.
A phenomenological parametrization for  nucleon to meson TDAs should properly incorporate
the polynomiality property of the Mellin moments of TDAs in $x_i$ as well as the support properties
described above.

In our approach we use the relation between GPDs and DDs - which is known to be a particular case of the Radon transform
\cite{Teryaev:2001qm} -
as the building block for the spectral representation of the multipartonic generalization of GPDs with the
restricted support properties. Indeed, representing GPDs as the Radon transform is the natural way to
implement polynomiality property known as the Cavalieri conditions in the Radon transform theory and
ensure the proper support in the longitudinal momentum fractions.
In order to write down the spectral representation for $\pi N$ TDAs
we introduce for each of  momentum fractions $x_i$ the spectral parameters
$\beta_{i}$, $\alpha_{i}$.
$x_{i}$
are supposed to have the following decomposition in terms of these spectral parameters:
$x_i=\xi+\beta_i+\alpha_i \xi$.
In order to satisfy the constraint
$x_1+x_2+x_3=2 \xi$
we  require that
$ \sum_i \beta_i=0$
and
$\sum_i  \alpha_i=-1$.

This allows to write down
the following spectral representation for $\pi N$ TDAs:
\be
&&
\! \! \! \! \!  \! \! \! \! \! H(x_1,\,x_2,\,x_3,\xi)|_{\sum x_i=2\xi}  \! 
= \!
\Big[
\prod_{i=1}^3
\int_{\Omega_i} d \beta_i d \alpha_i
\Big]
\delta(x_1-\xi-\beta_1-\alpha_1 \xi) \,
\delta(x_2-\xi-\beta_2-\alpha_2 \xi) \,
\nonumber \\ && \! \! \! \! \!  \! \! \! \! \!
\times
\delta(\beta_1+ \beta_2+ \beta_3)
\delta(\alpha_1+\alpha_2+\alpha_3+1)
 F(\beta_1, \, \beta_2, \, \beta_3, \, \alpha_1, \, \alpha_2, \alpha_3)\,.
\label{Spectral_for_GPDs_x123}
\ee
By
$\Omega_{i}$
we denote the copies of the usual DD square support in the parameter space:
$\Omega_{i}= \{ |\beta_{i}| \le 1\,; \ \ |\alpha_{i}| \le 1-|\beta_{i}|  \}$.
$F(\beta_1, \, \beta_2, \, \beta_3, \, \alpha_1, \, \alpha_2, \alpha_3)$
depends on six spectral parameters which are however subject to  constrains due
to two last $\delta$ function in (\ref{Spectral_for_GPDs_x123}).
Hence $F$ is  a quadruple distribution.
After introducing the natural combinations of spectral parameters and switching
to quark-diquark coordinates one may perform two integrations with the help of
$\delta$ functions and obtain the following spectral representation for
$\pi N$ TDAs:
\be
&&
\! \! \! \! \! H
(w,\,v,\,\xi)
\nonumber \\ && 
\! \! \! = \int_{-1}^1 d \sigma
\int_{-1+ |\sigma|/2}^{1- |\sigma|/2} d \rho
\int_{-1+|\sigma|}^{1-|\rho- \sigma/2 |-|\rho+ \sigma/2|} d \omega
\int_{-1/2+|\rho-  \sigma/2|+\omega/2}^{1/2-|\rho+ \sigma/2|-\omega/2} d \nu
\delta(w-\sigma-\omega \xi)
\nonumber \\ &&
\! \! \! \times  \delta(v-\rho-\nu \xi) \,
F (\sigma,\, \rho,\, \omega,\, \nu)\,,
\label{spectral_representation_D1}
\ee
where
$
F (\sigma,\, \rho,\, \omega,\, \nu) \equiv
F(\rho-\sigma/2,\, -\rho-\sigma/2,\, \sigma ,\, \nu - (1+\omega)/2,\, -\nu- (1+\omega)/2, \omega)
$
is the quadruple distribution.
One may check that the resulting $\pi N$ TDA possesses the desired support properties in $(w,v)$ variables.
The polynomiality property of the Mellin moments in $(w,v)$ is ensured by the very construction of
the spectral representation (\ref{spectral_representation_D1}).




\section{Radyushkin type Ansatz for $\pi N$ TDAs}

Now we discuss a possible approach for modelling quadruple distributions.
In analogy with GPDs \cite{RDDA4}
we suggest to employ the following factorized Ansatz:
\be
F(\sigma, \, \rho, \, \omega,\,\nu)=f(\sigma,\,\rho) \, h(\sigma,\,\rho,\,\omega,\,\nu)\,,
\label{Factorized_Ansatz}
\ee
where
$h(\sigma,\,\rho,\,\omega,\,\nu)$
is a profile function and
$f(\sigma,\rho)$
determines the shape of $\pi N$ TDA in the limit $\xi \rightarrow 0$.
Exploiting further the analogy with the usual GPDs we assume that the
shape of the profile
$h$
in
$(\omega, \nu)$
directions is determined by the asymptotic form of nucleon
distribution amplitude $\Phi^{\rm as}(y_1,y_2,y_3)=15/4 \, y_1 y_2 y_3$.
This results in the following simple expression for the profile in terms of the
initial constrained spectral parameters $\alpha_i$ and $\beta_i$:
\be
h(\beta_1, \beta_2, \beta_3\,; \alpha_1, \alpha_2, \alpha_3)
\Big|_{ \sum_i \beta_i =0 \atop \sum_i \alpha_i=-1}
=
\frac{15}{4}
\frac{ \prod_{i=1}^3 (1+\alpha_i-|\beta_i|) }{(1-\frac{1}{2} (|\beta_1|+|\beta_2|+|\beta_3|))^5}
\Big|_{ \sum_i \beta_i =0 \atop \sum_i \alpha_i=-1}
\,.
\label{profile}
\ee

For the moment we lack the knowledge of what may be the shape of $\pi N$ TDAs in the unphysical limit
$\xi \rightarrow 0$.
In the numerical exercises of \cite{Pire:2010if} we used a rather simple toy Ansatz
$
\left.f(\beta_1,\,\beta_2,\,\beta_3)\right|_{ \sum_i \beta_i =0}=
\left. 40/47 \, \prod_{i=1}^3 \theta(|\beta_i|\le 1) (1-\beta_i^2)\right|_{ \sum_i \beta_i =0}
$
normalized to unity. This form ensures the good convergence of the relevant integrals expressing
$\pi N$ TDAs in the spectral representation (\ref{spectral_representation_D1}).
A more realistic model should take into account that in
the limit $\xi \rightarrow 1$ $\pi N$ TDAs are fixed due to the soft pion theorem
\cite{Pobylitsa:2001cz}
and are expressed through
the nucleon DAs. In principle this allows to tune the shape of
$f(\sigma, \rho)$.
The recent calculations
\cite{Pasquini:2009ki}
of
$\pi N$
TDAs in the meson cloud model may also be useful for establishing the overall normalization.
It is worth to mention that employing the factorized Ansatz
(\ref{Factorized_Ansatz}) with the profile function (\ref{profile}) results in
the reliable methods for the calculation of the convolution integrals in $x_i$ occurring
for $\gamma^*N \rightarrow \pi N$  amplitude.

\section{Conclusions}
We  introduced the notion of quadruple distributions and construct
the spectral representation of TDAs involving
three parton correlators which arise in the description of baryon to meson
transitions.
We generalize Radyushkin's factorized Ansatz for the case of quadruple distributions and provide
the explicit expression for the corresponding profile function.
Our construction opens the way to quantitative modelling of baryon to meson TDAs
in their complete domain of  definition.
The main part of our analysis
can be directly applied to the nucleon to photon TDAs \cite{PS phys rev}. However the modelling
of the corresponding quadruple distributions has to account for the anomalous nature
of a photon \cite{Witten}. This is a nontrivial task which deserves separate studies.


We are   thankful to I.~Anikin, J.-Ph.~Lansberg, A.~Radyushkin and
S.~Wallon for many discussions and helpful comments.
K.S. also acknowledges  the partial support by Consortium Physique des Deux Infinis (P2I).
This work was supported by the Polish Grant N202 249235.



\bibliographystyle{aipproc}   

\begin{thebibliography}{9}

\bibitem{Frankfurt:1999fp}
  L.~L.~Frankfurt, P.~V.~Pobylitsa, M.~V.~Polyakov and M.~Strikman,
  \emph{Phys. Rev.  D} {\bf 60}, 014010 (1999)
  [arXiv:hep-ph/9901429].

\bibitem{Frankfurt:2002kz}
  L.~Frankfurt, M.~V.~Polyakov, M.~Strikman, D.~Zhalov and M.~Zhalov,
  ``Novel hard semiexclusive processes and color singlet clusters in hadrons,''
  in \emph{Newport News 2002, Exclusive Processes at High Momentum Transfer},
  edited by A.~Radyushkin and P.~Stoler;  World Scientific, Singapore, 2002, pp.361-368
    [arXiv:hep-ph/0211263].

\bibitem{Pire:2005ax}
  B.~Pire and L.~Szymanowski,
  \emph{Phys. Lett.  B}  {\bf 622}, 83 (2005)
  [arXiv:hep-ph/0504255].

\bibitem{Lansberg:2007ec}
  J.~P.~Lansberg, B.~Pire and L.~Szymanowski,
  \emph{Phys. Rev.  D} {\bf 75}, 074004 (2007)
  [Erratum-  \emph{ibid. D} {\bf 77}, 019902 (2008)]
  [arXiv:hep-ph/0701125];
\emph{Phys. Rev.  D} {\bf 76}, 111502
(2007)[arXiv:0710.1267 [hep-ph]].





\bibitem{RDDA1}
  A.~V.~Radyushkin,
  \emph{Phys. Rev.  D} {\bf 56}, 5524 (1997)
  [arXiv:hep-ph/9704207].






	

%
\bibitem{Radyushkin:1983ea}
  A.~V.~Radyushkin,
  \emph{Theor. Math. Phys.}  {\bf 61}, 1144 (1985)
  [\emph{Teor. Mat. Fiz.}  {\bf 61}, 284 (1984)];

\bibitem{RDDA4}
  I.~V.~Musatov and A.~V.~Radyushkin,
  \emph{Phys. Rev.  D} {\bf 61}, 074027 (2000)
  [arXiv:hep-ph/9905376].

\bibitem{Pire:2010if}
  B.~Pire, K.~Semenov-Tian-Shansky and L.~Szymanowski,
 \emph{Phys. Rev.  D} {\bf 82}, 094030 (2010)
  [arXiv:1008.0721 [hep-ph]].


\bibitem{Teryaev:2001qm}
  O.~V.~Teryaev,
  \emph{Phys. Lett.  B} {\bf 510}, 125 (2001)
  [arXiv:hep-ph/0102303].

\bibitem{Pasquini:2009ki}
  B.~Pasquini, M.~Pincetti and S.~Boffi,
  \emph{Phys. Rev.  D} {\bf 80}, 014017 (2009)
  [arXiv:0905.4018 [hep-ph]].

\bibitem{PS phys rev}
  B.~Pire and L.~Szymanowski,
  \emph{Phys. Rev.  D} {\bf 71}, 111501 (2005)
  [arXiv:hep-ph/0411387].

\bibitem{Pobylitsa:2001cz}
  P.~V.~Pobylitsa, M.~V.~Polyakov and M.~Strikman,
  \emph{Phys. Rev. Lett.} {\bf 87}, 022001 (2001)
  [arXiv:hep-ph/0101279].


\bibitem{Witten}
E.~Witten, \emph{Nucl. Phys. B} {\bf 120}, 189 (1977).


\end{thebibliography}


\end{document}